\magnification \magstep1
\raggedbottom
\openup 1\jot
\voffset6truemm
\def\cstok#1{\leavevmode\thinspace\hbox{\vrule\vtop{\vbox{\hrule\kern1pt
\hbox{\vphantom{\tt/}\thinspace{\tt#1}\thinspace}}
\kern1pt\hrule}\vrule}\thinspace}
\centerline {\bf NEW SUPPLEMENTARY CONDITIONS FOR A NON-LINEAR}
\centerline {\bf FIELD THEORY: GENERAL RELATIVITY}
\vskip 1cm
\leftline {Giampiero Esposito and Cosimo Stornaiolo}
\vskip 0.3cm
\noindent
{\it Istituto Nazionale di Fisica Nucleare, Sezione di Napoli,
Mostra d'Oltremare Padiglione 20, 80125 Napoli, Italy}
\vskip 0.3cm
\noindent
{\it Universit\`a degli Studi di Napoli Federico II, Dipartimento
di Scienze Fisiche, Complesso Universitario di Monte S. Angelo,
Via Cintia Edificio G, 80126 Napoli, Italy}
\vskip 1cm
\noindent
{\bf Abstract.}
The Einstein theory of general relativity provides a peculiar example
of classical field theory ruled by non-linear partial differential
equations. A number of supplementary conditions (more frequently
called gauge conditions) have also been considered in the
literature. In the present paper, starting from the de Donder gauge,
which is not conformally invariant but is the gravitational
counterpart of the Lorenz gauge, we consider, led by geometric
structures on vector bundles, a new family of gauges in general
relativity, which involve fifth-order covariant derivatives of
metric perturbations. A review of recent results by the authors
is presented: restrictions on the general form of the metric on
the vector bundle of symmetric rank-two tensor fields over
space-time; admissibility of such gauges in the case of linearized
theory about flat Euclidean space; generalization to a suitable
class of curved Riemannian backgrounds, by solving an integral
equation. Eventually, the applications to Euclidean quantum gravity
are discussed.
\vskip 100cm
\noindent
In the analysis of classical field theories which rely on partial
differential equations, the Einstein theory of general relativity
has a distinctive feature because it describes the gravitational
field as a non-linear system even in the absence of other 
fields.$^{1}$ The self-interaction of the gravitational field 
occurs because the space-time over which it propagates is defined
by gravity itself. Solutions of the 
Einstein field equations can be unique
only up to a diffeomorphism, and a fixed background metric is
introduced to obtain a definite member of the equivalence class
of metrics which represents a space-time. For this purpose, one has
also to impose four supplementary conditions on the covariant
derivatives of the physical metric with respect to the background
metric. The four degrees of freedom to make diffeomorphisms are
hence removed, and a unique solution for the metric components is
obtained. Moreover, since the metric defines the space-time
structure, one does not know in advance what the region is on 
which the solution should be determined.$^{1}$ All what one has is
a three-manifold $\Sigma$ with certain initial data $\cal I$ on it,
and one has to find a four-manifold $M$, an imbedding
$$
\theta: \Sigma \rightarrow M
$$
and a metric $g$ on $M$ which satisfies the Einstein equations
($T_{ab}$ denotes the energy-momentum tensor)
$$
R_{ab}-{1\over 2}g_{ab}R=8\pi G T_{ab},
\eqno (1)
$$
agrees with the initial values on $\theta(\Sigma)$, and is such
that $\theta(\Sigma)$ is a Cauchy surface for $M$.

On the other hand, 
the transformation properties of classical and quantum field 
theories under conformal rescalings of the metric
have led, over the years, to many deep developments in 
mathematics and theoretical physics, e.g. conformal-infinity
techniques in general relativity, twistor methods for gravitation
and Yang--Mills theory, the conformal-variation method in 
heat-kernel asymptotics, the discovery of conformal anomalies 
in quantum field theory. All these topics are quite relevant for 
the analysis of theories which possess a gauge freedom. As a first
example, one may consider the simplest gauge theory, i.e. vacuum
Maxwell theory in four dimensions in the absence of sources.
At the classical level, the operator acting on the potential $A_{b}$
is found to be
$$
P_{a}^{\; b}=-\delta_{a}^{\; b}\cstok{\ }
+R_{a}^{\; b}+\nabla_{a}\nabla^{b},
\eqno (2)
$$
where $\nabla$ is the Levi--Civita connection on space-time,
$\cstok{\ } \equiv g^{ab} \nabla_{a} \nabla_{b}$, and $R^{ab}$
is the Ricci tensor. Thus, the supplementary (or gauge) condition
of the Lorenz type, i.e.
$$
\nabla^{b}A_{b}=0
\eqno (3.a)
$$
is of crucial importance to obtain a wave equation for $A_{b}$.
The drawback of Eq. (3.{\it a}), however, is that it is not preserved
under conformal rescalings of the metric:
$$
{\widehat g}_{ab}=\Omega^{2} g_{ab}, \; \; \; \; 
{\widehat g}^{ab}=\Omega^{-2} g^{ab},
\eqno (4)
$$
whereas the Maxwell equations
$$
\nabla^{b}F_{ab}=0
\eqno (5)
$$
are invariant under the rescalings (4). This remark was the
starting point of the investigation 
by Eastwood and Singer,$^{2}$ who
found that a conformally invariant supplementary condition
may be imposed, i.e.
$$
\nabla_{b} \left[\Bigr(\nabla^{b}\nabla^{a}-2R^{ab}
+{2\over 3}R g^{ab} \Bigr)A_{a} \right]=0.
\eqno (6.a)
$$
As is clear from Eq. (6.{\it a}), conformal invariance is achieved at the
price of introducing third-order derivatives of the potential.
In flat backgrounds, such a condition reduces to
$$
\cstok{\ }\nabla^{b}A_{b}=0.
\eqno (7)
$$
Of course, all solutions of the Lorenz gauge are also solutions
of Eq. (7), whereas the converse does not hold.

Leaving aside the severe technical problems resulting from the attempt 
to quantize in the Eastwood--Singer gauge,$^{3}$ we are now interested
in understanding the key features of the counterpart for Einstein's
theory of general relativity. In other words, although the vacuum
Einstein equations
$$
R_{ab}-{1\over 2}g_{ab}R=0
\eqno (8)
$$
are not invariant under the conformal rescalings (4), we would like
to see whether the geometric structures 
leading to Eq. (6.{\it a}) admit
a non-trivial generalization to Einstein's theory, so that a
conformally invariant supplementary condition with a higher order
operator may be found as well. For this purpose, we re-express 
Eqs. (3.{\it a}) and (6.{\it a}) in the form
$$
g^{ab}\nabla_{a}A_{b}=0,
\eqno (3.b)
$$
$$ 
\eqalignno{
\; & g^{ab}\nabla_{a}\nabla_{b}\nabla^{c}A_{c}
+\left[\nabla_{b}\Bigr(-2R^{ba}+{2\over 3}R g^{ba} \Bigr)
\right]A_{a} \cr
&+ \left(-2R^{ba}+{2\over 3}R g^{ba} \right)
\nabla_{b}A_{a}=0.
&(6.b)\cr}
$$
Eq. (3.{\it b}) involves the space-time metric in its contravariant
form, which is also the metric on the bundle of 1-forms on $M$.
In Einstein's theory, one deals instead with the vector bundle of 
symmetric rank-two tensors on space-time
with DeWitt supermetric
$$
E^{abcd} \equiv {1\over 2} \Bigr(g^{ac}g^{bd}
+g^{ad}g^{bc}+ \alpha g^{ab}g^{cd} \Bigr),
\eqno (9)
$$
$\alpha$ being a real parameter different from $-{2\over m}$,
where $m$ is the dimension of space-time (this restriction on
$\alpha$ is necessary to make sure that the metric $E^{abcd}$
has an inverse). One is thus led to replace
Eq. (3.{\it b}) with the de Donder gauge
$$
W^{a} \equiv E^{abcd}\nabla_{b}h_{cd}=0.
\eqno (10)
$$
Hereafter, $h_{ab}$ denotes metric perturbations, since we are
interested in linearized general relativity.
The supplementary condition (10) is not invariant under conformal
rescalings, but the expression of the Eastwood--Singer gauge in
the form (6.{\it b}) suggests considering as a `candidate' for a
conformally invariant gauge involving a higher-order operator
the equation
$$
E^{abcd}\nabla_{a}\nabla_{b}\nabla_{c}\nabla_{d}W^{e}
+\left[\Bigr(\nabla_{p}T^{pebc}\Bigr)+T^{pebc}\nabla_{p}
\right]h_{bc}=0 .
\eqno (11)
$$
More precisely, Eq. (11) is obtained from Eq. (6.{\it b}) by applying
the replacement prescriptions 
$$
g^{ab} \rightarrow E^{abcd},
$$
$$
A_{b} \rightarrow h_{ab}, 
$$
$$
\nabla^{b}A_{b} \rightarrow W^{e},
$$ 
with $T^{pebc}$ a rank-four tensor field obtained from
the Riemann tensor, the Ricci tensor, the trace of Ricci and
the metric. In other words, $T^{pebc}$ is expected to involve
all possible contributions of the kind $R^{pebc}, R^{pe} g^{bc},
R g^{pe} g^{bc}$, assuming that it should be linear in 
the curvature.

When a supplementary (or gauge) condition is imposed in a theory
with gauge freedom, one of the first problems is to make sure that
such a condition is preserved under the action of the gauge 
symmetry. More precisely, either the gauge is originally satisfied,
and hence also the gauge-equivalent field configuration should fulfill
the condition, or the gauge is not originally satisfied, but one wants 
to prove that, after performing a gauge transformation, it is always
possible to fulfill the supplementary condition, eventually. The
latter problem is the most general, and has a well known counterpart
already for Maxwell theory. For linearized classical 
general relativity in the family of gauges described by Eq. (11),
the gauge symmetry remains the request of invariance under infinitesimal
diffeomorphisms. Their effect on metric perturbations is given by
$$
{ }^{\varphi}h_{ab} \equiv h_{ab}+(L_{\varphi}h)_{ab}
=h_{ab}+\nabla_{(a} \; \varphi_{b)}.
\eqno (12)
$$
For some smooth metric perturbation one might indeed have
(cf. Eq. (11))
$$
E^{abcd}\nabla_{a}\nabla_{b}\nabla_{c}\nabla_{d}W^{e}(h)
+\Bigr[(\nabla_{p}T^{pebc})+T^{pebc}\nabla_{p}\Bigr]h_{bc}
\not = 0.
\eqno (13)
$$
It is necessary to prove that one can, nevertheless, 
achieve the condition
$$
E^{abcd}\nabla_{a}\nabla_{b}\nabla_{c}\nabla_{d}
W^{e}({ }^{\varphi}h)+\Bigr[(\nabla_{p}T^{pebc})
+T^{pebc}\nabla_{p}\Bigr]{ }^{\varphi}h_{bc}=0.
\eqno (14)
$$
Equation (14) is conveniently re-expressed in a form where the
left-hand side involves a differential operator acting on the
1-form $\varphi_{q}$, and the right-hand side depends only on
metric perturbations, their covariant derivatives and the Riemann
curvature. Explicitly, one finds
$$
P_{e}^{\; q} \; \varphi_{q}=-F_{e},
\eqno (15)
$$
where (hereafter $h$ is the trace $g^{ab}h_{ab}$)
$$ \eqalignno{
\; & P_{e}^{\; q} \equiv \left(\nabla^{(c} \; \nabla^{d)}
\nabla_{c}\nabla_{d}+{\alpha \over 2} \cstok{\ }^{2}\right)
\left(\delta_{e}^{\; q} \cstok{\ }+\nabla^{q}\nabla_{e}
+\alpha \nabla_{e}\nabla^{q}\right) \cr
&+2 T_{\; \; e \; \; \; \; \; ;p}^{p \; (bq)} \;
\nabla_{b}+2T_{\; \; e}^{p \; (bq)} \nabla_{p}\nabla_{b},
&(16)\cr}
$$
$$ \eqalignno{
\; & F_{e} \equiv 2 \left(\nabla^{(c} \; \nabla^{d)}
\nabla_{c}\nabla_{d}+{\alpha \over 2}\cstok{\ }^{2}\right)
\left(\nabla^{q}h_{qe}+{\alpha \over 2}\nabla_{e}h \right) \cr
&+2 T_{\; \; e \; \; \; \; ;p}^{p \; \; bc} \; h_{bc}
+2 T_{\; \; e}^{p \; \; bc} \; \nabla_{p}h_{bc}.
&(17)\cr}
$$

Our original work in Ref. 4 has proved the following results:
\vskip 0.3cm
\noindent
(i) The value $\alpha=-2$ in the DeWitt supermetric (9) is
ruled out if one wants to be able to solve Eq. (15) for
$\varphi_{q}$. 
\vskip 0.3cm
\noindent
(ii) If $\alpha=-1$, the general solution of Eq. (15) in
flat $m$-dimensional Euclidean space ${\bf E}^{m}$ reads
$$ \eqalignno{
\; & \varphi_{a}(x)=\Omega_{a}(x)+\int_{{\bf E}^{m}}
G_{a}^{\; b}(x,y)w_{b}(y)dy \cr
&+\int_{{\bf E}^{m}} \int_{{\bf E}^{m}}
G_{a}^{\; b}(x,y) G_{b}^{\; c}(y,z)v_{c}(z)dy \; dz \cr
&+2(2\pi)^{-{m\over 2}}\int_{{\bf E}^{m}}
|\xi|^{-6} {\widetilde F}_{a}(\xi)
e^{i \xi \cdot x} d\xi,
&(18)\cr}
$$
where $\Omega_{a},w_{a}$ and $v_{a}$ are harmonic 1-forms 
in ${\bf E}^{m}$, $G_{a}^{\; b}$ is the Green kernel of the
Laplacian acting on 1-forms, and $|\xi| \equiv 
\sqrt{\xi_{a}\xi^{a}}$. In the last integral in Eq. (18) the
only poles of the integrand occur when 
$$
\xi_{0}=\pm i \sqrt{ \sum_{k=1}^{m-1}\xi_{k}\xi^{k}},
$$
i.e. on the imaginary $\xi_{0}$ axis. Thus, integration on the
real line for $\xi_{0}$, and subsequent integration with respect
to $\xi_{1},...,\xi_{m-1}$, yields a well defined integral
representation of $\varphi_{q}$. 
\vskip 0.3cm
\noindent
(iii) On compact Riemannian manifolds $(M,g)$ without boundary
and with non-vanishing Riemann curvature, Eq. (15) can be turned
into the integral equation
$$ \eqalignno{
\; & \varphi_{e}(x)+\int_{M}{\cal G}_{e}^{\; p}(x,y)
\Bigr({\cal B}_{p}^{\; r}\varphi_{r}\Bigr)(y)
\sqrt{{\rm det} \; g(y)} \; dy \cr
&+\int_{M}{\cal G}_{e}^{p}(x,y)F_{p}(y)
\sqrt{{\rm det} \;g(y)} \; dy=0,
&(19)\cr}
$$
where ${\cal G}_{e}^{\; p}$ is the Green kernel of the operator 
$$
{\cal A}_{e}^{\; q} \equiv \left(\nabla^{(c} \; \nabla^{d)}
\nabla_{c}\nabla_{d}+{\alpha \over 2} \cstok{\ }^{2}\right)
\left(\delta_{e}^{\; q} \cstok{\ }+\nabla^{q}\nabla_{e}
+\alpha \nabla_{e}\nabla^{q}\right), 
\eqno (20)
$$
and
$$
{\cal B}_{e}^{\; q} \equiv
2 T_{\; \; e \; \; \; \; \; ;p}^{p \; (bq)} \; \nabla_{b}
+2T_{\; \; e}^{p \; (bq)} \nabla_{p}\nabla_{b}.
\eqno (21)
$$
A recursive algorithm for the solution of Eq. (19) can be 
developed provided that ${\cal B}_{e}^{\; q}$ is a symmetric
elliptic operator, so that it admits a discrete spectral
resolution with eigenvectors of class $C^{\infty}$. The 
ellipticity condition means that the leading symbol of
${\cal B}_{e}^{\; q}$ is non-vanishing for $\xi \not =0$, i.e.
$$
-2T_{\; \; e}^{p \; (bq)} \xi_{p}\xi_{b} \not = 0
\; {\rm for} \; \xi \not = 0.
\eqno (22)
$$
This condition receives contributions from the parts of $T$ involving
the Ricci tensor and the scalar curvature, but not from the Riemann
tensor, which is antisymmetric in $b$ and $q$. For a given choice
of background with associated curvature and tensor $T$, the above
condition provides a useful operational criterion to check the
admissibility of our supplementary condition (11). 

The unsolved problem of our
investigation is how to choose, or determine, the form of the
tensor field $T^{pebc}$ in the supplementary condition (11). If
one writes for $T^{pebc}$ the most general combination of Riemann,
Ricci, trace of Ricci and background metric, 
it remains very difficult to study the behaviour of Eq. (11) 
under conformal rescalings. For example, the term which is known
explicitly in Eq. (11) reads
$$
\eqalignno{
\; & E^{abcd}\nabla_{a}\nabla_{b}\nabla_{c}\nabla_{d}W^{e}
=(2+\alpha)\cstok{\ }^{2}W^{e} \cr
&+2 \left[\left(\Bigr(\nabla_{h}R^{de}\Bigr)
-\Bigr(\nabla^{e}R_{\; h}^{d}\Bigr)\right)
\Bigr(\nabla_{d}W^{h}\Bigr)
+\Bigr(\nabla^{h}R \Bigr)\Bigr(\nabla_{h}W^{e}\Bigr)\right] \cr
&+2R^{ah}\nabla_{a}\nabla_{h}W^{e}
+{3\over 2}R_{\; qab}^{e} \; R_{h}^{\; qab} \; W^{h},
&(23)\cr}
$$
and one has, under conformal rescalings, the well known 
transformation properties of Riemann and Ricci, jointly with
$$
{\widehat W}^{a}=\Omega^{-4}\left[W^{a}+(m-2)
h^{ar}Y_{r}-(1+\alpha){\widehat h} g^{ar}Y_{r} \right],
\eqno (24)
$$
where $Y_{r} \equiv \nabla_{r} \log \Omega$,
${\widehat h} \equiv g^{cd}h_{cd}$. The next task is to check 
whether the conformal variation of the right-hand side 
of (23) compensates the conformal
variation of $\nabla_{p}(T^{pebc}h_{bc})$ for a suitable
form of $T^{pebc}$. It should also be stressed that the results
(ii) and (iii) deal with the Riemannian rather than the Lorentzian 
case. The work in Ref. 4 has also performed the analysis in a
Minkowskian background, but a curved Lorentzian background might
lead to some novel features because it is then impossible to use
the spectral theory of elliptic operators on manifolds.

The above results and open problems seem to suggest that new 
perspectives are in sight in the investigation of supplementary
conditions in general relativity. They might have applications both
in classical theory (linearized equations in gravitational wave 
theory, symmetry principles and their impact on gauge conditions),
and in the attempts to quantize the gravitational field. In 
particular, the quantization via Euclidean path integrals requires
adding to the Euclidean Einstein-Hilbert action $I_{EH}$ 
(supplemented by a boundary term) the integral $I_{GA}$ over
$M$ of $\chi^{a}\beta_{ab}\chi^{b}$, where $\chi^{a}$ is a 
gauge-averaging functional$^{5}$ 
and $\beta_{ab}$ is an invertible matrix.
The sum of the integrals $I_{EH}$ and $I_{GA}$ is what we mean by
full Euclidean action for gravity (but there is, of course, also
the ghost action.$^{5}$)
If $\chi^{a}$ contains fifth-order covariant derivatives of $h_{ab}$
and curvature terms as we have proposed, it is not {\it a priori}
obvious that the full Euclidean action remains unbounded from 
below.$^{6}$ One might instead hope to combine ellipticity of the
theory (now ruled by the leading symbol of a tenth-order differential
operator resulting from $\chi^{a}\beta_{ab}\chi^{b}$) with the need
to obtain a full Euclidean action for gravity which is bounded 
from below. For this purpose, only explicit calculations with a
definite form of the tensor $T^{pebc}$ can help to settle the issue.
\vskip 0.3cm
\leftline {\bf Acknowledgments}
\vskip 0.3cm
\noindent
This work has been partially supported by PRIN97 `Sintesi'. 
\vskip 0.3cm
\leftline {\bf References}
\vskip 0.3cm
\item {1.}
S. W. Hawking and G. F. R. Ellis, {\it The Large-Scale Structure
of Space-Time} (Cambridge University Press, Cambridge, 1973).
\item {2.}
M. Eastwood and I. M. Singer, {\it Phys. Lett.} 
{\bf A107}, 73 (1985).
\item {3.}
G. Esposito, {\it Phys. Rev.} {\bf D56}, 2442 (1997).
\item {4.}
G. Esposito and C. Stornaiolo, `A New Family of Gauges in
Linearized General Relativity' (GR-QC 9812044).
\item {5.}
B. S. DeWitt, in {\it General Relativity, an Einstein Centenary
Survey}, eds. S. W. Hawking and W. Israel (Cambridge University
Press, Cambridge, 1979).
\item {6.} 
S. W. Hawking, in {\it General Relativity, an Einstein Centenary
Survey}, eds. S. W. Hawking and W. Israel (Cambridge University
Press, Cambridge, 1979).

\bye